# A Linear Classifier Based on Entity Recognition Tools and a Statistical Approach to Method Extraction in the Protein-Protein Interaction Literature

Anália Lourenço[1], Michael Conover[2,3], Andrew Wong[4], Azadeh Nematzadeh[2,3], Fengxia Pan[5*], Hagit Shatkay[4,6,7§], Luis M. Rocha[2,3§]

[1] Institute for Biotechnology and Bioengineering, Centre of Biological Engineering, University of Minho, Braga, Portugal
[2] School of Informatics and Computing, Indiana University, Bloomington, IN, USA
[3] FLAD Computational Biology Collaboratorium, Instituto Gulbenkian de Ciência, Portugal
[4] School of Computing, Queen's University, Kingston, ON, Canada
[5] Microsoft Corp., Redmond, WA, USA
[6] Dept. of Computer and Information Sciences, University of Delaware, Newark, DE, USA
[7] Center for Bioinformatics and Computational Biology, Delaware Biotechnology Institute, University of Delaware, Newark, DE, USA

[§]Corresponding author
[*]Fengxia Pan's contribution to the work was performed while she was a graduate student at Queen's University, Kingston, ON, Canada

Email addresses:
    AL: analia@deb.uminho.pt
    MC: midconov@indiana.edu
    AW: 3aw14@queensu.ca
    AN: azadnema@indiana.edu
    FP: fepan@microsoft.com
    HS: shatkay@cis.udel.edu
    LR: rocha@indiana.edu

# Abstract

**Background**

We participated, as Team 81, in the *Article Classification* and the *Interaction Method* subtasks (ACT and IMT, respectively) of the *Protein-Protein Interaction* task of the BioCreative III Challenge. For the ACT, we pursued an extensive testing of available Named Entity Recognition and dictionary tools, and used the most promising ones to extend our Variable Trigonometric Threshold linear classifier. Our main goal was to exploit the power of available named entity recognition and dictionary tools to aid in the classification of documents relevant to Protein-Protein Interaction (PPI). For the IMT, we focused on obtaining evidence in support of the interaction methods used, rather than on tagging the document with the method identifiers. We experimented with a primarily statistical approach, as opposed to employing a deeper natural language processing strategy. In a nutshell, we exploited classifiers, simple pattern matching for potential PPI methods within sentences, and ranking of candidate matches using statistical considerations. Finally, we also studied the benefits of integrating the method extraction approach that we have used for the IMT into the ACT pipeline.

**Results**

For the ACT, our linear article classifier leads to a ranking and classification performance significantly higher than all the reported submissions to the challenge in terms of *Area Under the Interpolated Precision and Recall Curve*, *Mathew's Correlation Coefficient*, and *F-Score*. We observe that the most useful Named Entity Recognition and Dictionary tools for classification of articles relevant to protein-protein interaction are: ABNER, NLPROT, OSCAR 3 and the PSI-MI ontology. For the IMT, our results are comparable to those of other systems, which took very different approaches. While the performance is not very high, we focus on providing evidence for potential interaction detection methods. A significant majority of the evidence sentences, as evaluated by independent annotators, are relevant to PPI detection methods.

**Conclusions**

For the ACT, we show that the use of named entity recognition tools leads to a substantial improvement in the ranking and classification of articles relevant to protein-protein interaction. Thus, we show that our substantially expanded linear classifier is a very competitive classifier in this domain. Moreover, this classifier produces interpretable surfaces that can be understood as "rules" for human understanding of the classification. We also provide evidence supporting certain named entity recognition tools as beneficial for protein-interaction article classification, or demonstrating that some of the tools are not beneficial for the task. In terms of the IMT task, in contrast to other participants, our approach focused on identifying sentences that are likely to bear evidence for the application of a PPI detection method, rather than on classifying a document as relevant to a method. As BioCreative III did not perform an evaluation of the evidence provided by the system, we have conducted a separate assessment, where multiple independent annotators manually evaluated the evidence produced by one of our runs. Preliminary results from this experiment are reported here and suggest that the majority of the evaluators agree that our tool is indeed effective in detecting relevant evidence for PPI detection methods. Regarding the integration of both tasks, we note that the time required for running each pipeline is realistic within a curation effort, and that we can, without compromising the quality of the output, reduce the time necessary to extract entities from text for the ACT pipeline by pre-selecting candidate relevant text using the IMT pipeline.

# Background

A basic step toward discovering or extracting information about a particular topic in biomedical text, is the identification of a set of documents deemed relevant to that topic. Separating relevant from irrelevant documents is an example of *document classification*. Due to the central role document classification plays in biomedical literature mining, part of the *BioCreative* (Critical Assessment of Information Extraction systems in Biology) challenge evaluation is the *Article Classification Task* (ACT). In the last three challenges this task has focused on the classification of articles based on their relevance to Protein-Protein Interaction (PPI) [14].

For the BioCreative challenges 2 (BC2) and 2.5 (BC2.5) we have developed the lightweight *Variable Trigonometric Threshold* (VTT) linear classifier that employs word-pair textual features and protein counts extracted using the ABNER tool [20]. VTT was one of the top performing classifiers in the abstract classification task of BC2 [1] and the best classification system on the full-text scenario of BC2.5 [13] as tallied by the organizers [16].

In this BioCreative 3 challenge (BC3), we developed a novel and more general version of VTT which utilizes a number of features obtained via *Named Entity Recognition* (NER) and dictionary tools. We continue the development of this simple linear classifier since it has performed very well in the real-world scenarios of BioCreative, where training and test data are not guaranteed to be drawn from the same distributions of features; the simple linear decision surface seems to generalize the concept of PPI better than more sophisticated classifiers in this context [13]. We show that by expanding the classifier to handle a substantial increase in the amount of NER data, its performance improves significantly. Another interesting feature of the VTT is the interpretability of its simple decision surface, leading to (linear) "rules" for deciding the relevance of literature to PPI.

Throughout the development of our classifier, we analyzed the applicability of various NER and dictionary tools for deciding PPI-relevance. The assessment of appropriate tools is also described in this article, and offered to the community as a large-scale empirical study. In addition, we examine a few other questions related to the VTT and article classification. First, is there a benefit to using word bigrams as textual features, compared to the smaller set of word-pairs we previously employed [1, 13]? Second, does full-text data (when available) benefit classification? This last question is approached only partially here; as full-text data was not fully provided by BC3, we harvested a full-text subset for those BC3 articles that were available through *PubMed Central*.

The *Interaction Method Task* (IMT) at BC3, looked beyond the identification of relevant articles, and posed the challenge of finding evidence within full-text biomedical publications concerning the technique used for identifying protein-protein interaction. The task definition made the point that*: "A crucial aspect for the correct annotation of experimentally determined protein interactions is to determine the technique described in the article to support a given interaction… For this task, we will ask participants to provide, for each full text article, a ranked list of interaction detection methods, defined by their corresponding unique concept identifier from the PSI-MI ontology"* [14]. It also required including, as part of the submission for each Interaction Method, the evidence string derived from the text that supports the decision to associate the method with the article.

We thus literally interpreted the IMT task as that of finding, within the text, discussion of the *used techniques* that can be utilized for detecting PPIs, rather than that of identifying the PPIs themselves. Consequently, we took the approach of looking within the text for sentences that are likely to form *evidence for methods being employed*, tagging articles with the (likely) methods found. We then provided, in accordance with the BioCreative IMT output specification, for each article the identifiers of these methods, along with a score indicating the level of confidence our system associates with each method. This score reflects how confident the system is in making the association between the method and the article. The sentences within the text on which the association was based were provided as evidence.

Almost all teams participating in the BioCreative III IMT challenge, regarded the method-assignment as an article classification task, in which articles are assigned to one (or more) of the many different PPI methods as categories. In contrast, we have taken a very different route. We focused primarily on identifying *potential evidence for the use of methods* within the text, and then narrowed the candidate sentences to those who may discuss methods that *can be used* for PPI detection. Once sentences were found that were likely to bear evidence for the use of a potential PPI method, we scored these sentences with respect to the associated PPI detection method; PPI methods associated with high-scoring sentences were then listed as PPI methods supported by the article, with the high scoring sentences listed as evidence. Thus, the fundamental difference between our system and the other participating systems is that we focused on identifying evidence for potential use of PPI *detection methods*, while most other systems focused on classifying documents into method-categories, without searching for the explicit evidence.

Moreover, in contrast to other teams, which based their work on using natural language processing (NLP) to identify a variety of components and named entities, including proteins (Wang et al, Rinaldi et al, Matos et al) and possibly interactions among them (Rinaldi et al), as a fundamental step prior to method detection, we only used simple pattern matching of methods, ranking candidate matches using statistical considerations, without making an attempt at identifying entities. We do believe that NER to identify proteins is likely to improve our system's performance, but as said, we have focused on identification of methods that *can be used for identifying PPI,* rather than on the PPIs themselves.

Another notable aspect of IMT and its evaluation, is that while the task definition required associating methods with articles, providing the ranking and the strength of the association as well as the *evidence* supporting it, the evaluation only measured whether the correct method-identifiers were associated with each article, regardless of the strength assigned to this association, *and regardless of the evidence*. Correctness was determined by comparison of the method identifiers assigned by the system to the method identifiers assigned by human annotators. The evidence, which was requested in the task specification, was not formally evaluated or examined in BC3.

Furthermore, the training data consisted strictly of full text articles along with the PPI detection method tags assigned to the articles by curators, but did not provide any indication or tagging of the evidence within the text supporting this assignment. Similarly, the gold standard released after the challenge does not show this evidence. As such, there is currently no data against which one can evaluate the quality of the evidence produced by the competing systems.

To overcome this shortfall in both the data and the evaluation, immediately following the BioCreative meeting, we have recruited a team of independent annotators to go over the results produced from one of our runs, and constructed a triply-annotated corpus of over 1000 sentences. The section on the Interaction Methods Task, and its Results subsection, provide further detail about the use of this corpus in our evaluation.

## Article Classification Task

We participated in both the online (via the BioCreative MetaServer platform) and the offline components of ACT. We used four distinct versions of the most general VTT linear classifier as presented below. The main goal was to study the effect of using various NER and dictionary tools on classification performance. Therefore, the four versions of the VTT vary in the amount and the type of NER data which they use.

**Data and Feature Extraction**

**Training Corpora**

Given a labeled training corpus of documents *D*, let *P* refer to the set of documents labeled relevant or positive, and *N* to the set of documents labeled irrelevant or negative; by definition, $D \equiv P \cup N$ and $P \cap N \equiv \emptyset$. All documents, $d \in D$, are preprocessed by removal of stopwords[1] and Porter Stemming [18]. For training data we used the training and development sets released by BC3 for the ACT, as well as the documents released for IMT, which we labeled as positive. This results in a set of 8315 unique documents (3857 labeled positive, and 4458 labeled negative) defined by their PubMed IDs (PMID). To produce textual features (as described below), we oversampled documents from the positive set to obtain a balanced set where $|P| = |N| = 4458$, $|D| = 8916$. By oversampling we mean that we randomly selected positive documents to be repeated in the set *P*. For textual feature selection, as described below, we used only the title and abstract text associated with the PubMed records of these documents. For NER feature selection (see below), we extracted figure caption text and full text from the subset of public-domain documents with PubMed Central records. We denote the full text subset as: $D^{PMC} \subset D$, where $|D^{PMC}| = 4190$ (≈50% of *D*).

**Test/Validation Corpora**

Let $D_t$ refer to the official BC3 test set of documents, which was unlabeled at the time of the challenge, but whose class labels were subsequently provided to the community as a gold standard. This is a highly unbalanced set, with 5090 negative or irrelevant documents, and 910 positive or relevant documents, for a total of $|D_t| = 6000$ documents. Out of these, we were able to obtain PubMed Central records for $|D_t^{PMC}| = 3019$ documents (60% of $D_t$); 423 positives and 2596 negatives (preserving a similar proportion of negatives to positives as in the overall test set).

**Textual Feature Selection: Word-Pair and Bigram features**

The VTT classifier requires textual features to have been obtained from labeled, training documents. In previous versions of VTT, we have used word-pair features similar to bigrams, but which are less computationally demanding to obtain [1, 13]. Here, because we are interested in investigating the benefit of using our word-pairs compared with bigrams, we have used both types of features in different runs of the classifier.

The *Short-window word-Pair* features (denoted SP) are computed by first selecting the set of top 1000 words, *W*, obtained by ranking all words occurring in the balanced training corpus according to the following score:

$$S(w) = |p_P(w) - p_N(w)|, \text{ where } w \in W^D,$$

and $W^D$ is the set of all unique words in the training corpus *D*, after pre-processing and stopword removal. The score, $S(w)$, measures the difference between the probability of occurrence of a word *w* in relevant documents, $p_P(w)$, and the probability of occurrence in irrelevant ones, $p_N(w)$. Each document in the set *D* is subsequently converted into an ordered list comprised of a subset of these 1000 words, $w \in W$. The list representing each document is ordered (with repetition) according to the sequence in which the words occur in the original text. That is, the original sequence of words in the text, is converted into a sequence that contains only words $w \in W$; all words not in the top 1000 set, *W*, are removed. The top 10 (stemmed) words and their *S* score in the training data for BC3 were: `interact` (0.41), `protein` (0.4), `bind` (0.33), `domain` (0.27), `complex` (0.26), `regul` (0.24), `activ` (0.21), `here` (0.19), `phosphoryl` (0.16), `function` (0.15).

---

[1] The list of stop-words removed: *i, a, about, an, are, as, at, be, by, for, from, how, in, is, it, of, on, or, that, the, this, to, was, what, when, where, who, will, the, and, we, were*. Note that words "with" and "between" were kept.

From the ordered lists of words representing the documents, we extract the *SP* features $(w_i, w_j)$: pairs of consecutive words from the ordered lists that represent documents[2]. The order in which words occur is preserved, i.e. $(w_i, w_j) \neq (w_j, w_i)$. For each SP feature we compute its probability of occurring in a positive and in a negative document: $p_P(w_i, w_j)$ and $p_N(w_i, w_j)$, respectively. Figure 1 depicts the 1000 SP features with largest $S(w_i, w_j) = |p_P(w_i, w_j) - p_N(w_i, w_j)|$, plotted on a plane where the horizontal axis is the value $p_P(w_i, w_j)$ and the vertical axis is the value $p_N(w_i, w_j)$; we refer to this plane as the $p_P/p_N$ plane. Table 1 lists the top 10 SP features for score *S*.

The *Bigram* features are extracted very similarly, except that we compute the word-pair probabilities $p_P/p_N$ for *all* consecutive word-pair occurrences in the original text (after stemming and stop word removal), rather than restricting the pairs to the ordered list representation of documents as done for the SP features. Bigram feature extraction results in a much more computationally demanding process, because the set of observed bigrams is much larger than the set of observed SP word-pairs built from the fixed set of 1000 words. Table 2 lists the top 10 bigram features for score *S*, which are very similar to the top 10 SP features.

One side goal of this work was to investigate whether the computational overhead of bigram extraction is worthwhile. Notably, the generation of SP features requires two iterations over each document: one to extract the single word features, and another to obtain the occurrence counts of SP features after ranking of single word features over the entire training corpus. In contrast, bigrams in principle require a single iteration over each document to extract occurrences. However, there are many more unique observable bigrams than unique single word features, due to the possible combinations of single words with one another. In contrast, the second pass to compute SP features is not over the entire document text, but over the ordered lists containing only the top (1000) single words, which results in a much smaller set of possible word pairs. Therefore, in a large corpus the list of bigrams to store and index for tallying occurrences is much larger than that of SP features, resulting in a substantial computational overhead. One other possible issue is that of finding the optimal number of top scoring words selected to produce SP features. We showed in an earlier publication [1] that the *S* score histogram can guide us to identify a good threshold number after which no improvement results. We used this technique here.

For simplicity, in the remainder of the article, unless otherwise specified, we refer to textual features simply by the symbol *w*.

**Entity Count Features: data from Entity Recognition Tools**
In our previous work with a simpler version of VTT for BC2 [1], we used as an additional feature the *number of proteins mentioned* in abstracts, as identified by the NER tool ABNER [20]. More recently, in BC2.5, we used the same additional feature in distinct sections of full text documents, and observed that terms extracted from domain ontologies did not help in article classification [13]. Here, we pursue a much wider investigation of the utility of using terms from NER and dictionary tools available to the community.

---

[2] Note that the ordered lists representing documents contain only words from set *W* (1000 top words). Therefore, adjacent words in such a list may or may not be adjacent in the original text; we refer to these word-pairs as "short-window" pairs.

What we use for VTT are *entity count features*: for each document $d \in D$, we compute the number of occurrences $n_\pi(d)$, of each entity type $\pi$. An example of an entity type is "protein mentions" as identified by ABNER. Naturally, in the context of BC3, we are interested in the entity count features that can best discriminate documents relevant for PPI (positive) from irrelevant ones (negative). For that purpose, we utilized the NER tools ABNER [20, 21], NLProt [17] and OSCAR 3 [3, 8] and compiled dictionaries from the BRENDA (enzymes) [4, 19] and ChEBI (chemical compounds) [9] databases, as well as the PSI-MI ontology (experimental methods) [7].

With each one of these tools we extracted various types of entity count features in abstracts for all documents, $d \in D$, and also in figure captions and full text of the subset of documents available in PubMed Central, $d \in D^{PMC}$. Examples of entity count features we collected are the *number of protein mentions in an abstract identified by NLProt*, and the *number of PSI-MI method mentions in figure captions*.

Finally, we selected those entity feature counts that best discriminated relevant from irrelevant documents in the training data $D$ and $D^{PMC}$. The selection was done by computing the probability of finding, in the training data, positive and negative documents, $d$, with at least $x$ mentions of entity $\pi$: $p_P(n_\pi(d) \geq x)$ and $p_N(n_\pi(d) \geq x)$, respectively. The relationship between these quantities, for a given entity, is best appreciated in graphical form: Figure 2 depicts a comparison of these probabilities for *ABNER protein mentions in abstracts* of documents in $D$, and for *CHEBI compound names in full text documents* in $D^{PMC}$. As can be seen in this figure, the counts of CHEBI compound name mentions in full text documents are not very distinct for documents labeled positive or negative. In contrast, counts of ABNER protein mentions in abstracts are quite distinct for relevant and irrelevant documents; we can see, for instance, that 90% of all positive documents in $D$ have 5 or more protein mentions, whereas only 40% of negative documents have the same number of mentions.

We used this type of chart to identify which features from NER and dictionary tools behave differently for relevant and irrelevant documents. Specifically, we identified those entity count features for which the difference in occurrence probability, $|p_P(n_\pi(d) \geq x) - p_N(n_\pi(d) \geq x)|$, is greater than 0.3 for some $x$ number of mentions[3]. If the sign of the difference $(p_P(n_\pi(d) \geq x) - p_N(n_\pi(d) \geq x))$ is positive, we consider the entity count feature $\pi$ to be *positively correlated with the set of positive documents* (*P*) in the training data, and *positively correlated with the set of negative documents* (*N*) otherwise. Using this criterion on all the NER data we produced, we identified only 5 entity count features positively correlated with *P*, and none positively correlated with *N*:

1. ABNER protein mentions in abstracts
2. NLProt protein mentions in abstracts
3. OSCAR compounds in abstracts
4. ABNER protein mentions in figure captions
5. PSI-MI methods in full texts

The charts with the $p_P(n_\pi(d) \geq x)$ and $p_N(n_\pi(d) \geq x)$ probabilities are shown in Figures 2 (for ABNER in abstracts) and 3 (for the other 4 entity count features). Notice that entity count features 4 and 5 above are only defined for documents in the $D^{PMC}$ subset. We rejected all other entity count features according to the criteria above; these include all counts obtained via BRENDA and CHEBI. We provide the charts for all tested entity count features in supplementary materials[4].

---
[3] We observed that entity counts with values lower than 0.3 hindered performance of the VTT classifier.

**Approach: Variable Trigonometric Threshold Classifier**

We present here a more general, and novel, formulation of the VTT classifier, which can integrate information from various textual and entity count features. A document *d* is considered to be relevant if:

$$M \cdot \frac{P(d)}{N(d)} \geq \lambda_0 + \sum_{\pi=1}^{EP} \frac{\beta_\pi - n_\pi(d)}{\beta_\pi} - \sum_{\nu=1}^{EN} \frac{\beta_\nu - n_\nu(d)}{\beta_\nu}$$

Eq. (1)

and irrelevant otherwise. The above expression defines a linear decision surface for classifying documents. The left-hand side contains the sum of the contributions from textual features for a positive, *P(d)*, and a negative, *N(d)*, decision for document *d*, which are computed from the $p_P/p_N$ plane of textual features as:

$$P(d) = \sum_{w \in d} \cos(\alpha(w)) = \sum_{w \in d} \frac{p_P(w)}{\sqrt{p_P^2(w) + p_N^2(w)}}$$

$$N(d) = \sum_{w \in d} \sin(\alpha(w)) = \sum_{w \in d} \frac{p_N(w)}{\sqrt{p_P^2(w) + p_N^2(w)}}$$

Eq. (2)

where *w* denotes a textual feature such as SP or bigram as described above. In other words, *P(d)* sums the cosine contributions of every occurring feature *w* in document *d*, when projected on the $p_P/p_N$ plane. *N(d)*, in turn, sums the sine contributions of every occurring feature *w*.

The right-hand side expression of Eq. (1) specifies a *decision threshold* for a document, given its ratio of positive and negative textual feature contributions (on the left-hand side). This decision threshold is defined by a constant, $\lambda_0$, and a *variable* component, defined by entity count features. The idea is that information from NER data can alter the decision threshold. For instance, in Figure 2 we can see that 90% of all positive documents in the training data set, *D,* have 5 or more ABNER-extracted protein mentions, whereas only 40% of negative documents have the same number of mentions. Therefore, when a given document, *d,* contains more than 5 ABNER-extracted protein mentions, we can expect it to have a higher chance of being relevant. To introduce this type of information into the decision threshold, the VTT classifier is defined for *M=|EP-EN|* entity count features, *EP* of which are positively correlated with positive documents (such as ABNER protein mentions), and *EN* of which are positively correlated with negative documents. For simplification, we refer to the first as *positive entity count features*, and to the second as *negative entity count features*.

---

[4] http://cnets.indiana.edu/groups/casci/piare

Each positive entity count feature $\pi$ adjusts the decision threshold for document $d$ with the factor $(\beta_\pi - n_\pi(d))/\beta_\pi$, where $\beta_\pi$ is a constant parameter; when $n_\pi(d) > \beta_\pi$, the threshold is lowered, and increased otherwise. Each negative entity count feature $\nu$ adjusts the decision threshold for document $d$ with the factor $(n_\nu(d)-\beta_\nu)/\beta_\nu$, where $\beta_\nu$ is a constant; when $n_\nu(d) > \beta_\nu$, the threshold is increased, and lowered otherwise. The β parameters represent the *neutral threshold point* for the respective entity count feature: when $n_\pi(d) = \beta_\pi$, there is no threshold adjustment from information about entity count feature $\pi$.

It is easy to visualize the VTT linear decision surface, even with many different entity count features. We can plot the decision surface and every document $d$ in a plane, where the horizontal and vertical coordinates are defined as:

$$x(d) = \frac{P(d)}{N(d)}$$

$$y(d) = \frac{1}{M}\left(\sum_{\pi=1}^{EP}\frac{n_\pi(d)}{\beta_\pi} - \sum_{\nu=1}^{EN}\frac{n_\nu(d)}{\beta_\nu}\right),$$

Eq. (3)

where $M=|EP-EN|$. In this plane, the decision surface is simply given by:

$$y(d) = \lambda - x(d) \quad \text{Eq. (4)}$$

where λ is a constant ($\lambda = \lambda_0/M + 1$, but we treat it as a constant parameter to be searched, so the value of $\lambda_0$ is irrelevant). Figure 4 depicts the decision plane of VTT; negative documents are expected to plot near the origin and positive documents above the decision line. A few interesting points naturally derive from this plane. Given a document $d$, we compute the values of $x(d)$ and $y(d)$ according to Eq. (3). The decision is then calculated by comparing $x(d)$ with the decision threshold $T(d) = \lambda - y(d)$ given by Eq. (4); if $x(d) > T(d)$, $d$ is considered relevant, and irrelevant otherwise (see Figure 4, left). Therefore, a *measure of confidence* in the decision can be derived from the difference $\delta(d) = |x(d) - T(d)|$, which can be normalized by dividing it by the maximum value of δ in the training data $D$:

$$C(d) = \frac{\delta(d)}{\max(\delta(d), \forall_{d \in D})}$$

Eq. (5)

In addition to the class decision, computed by the VTT decision surface (Eq. (4)), we ranked positive documents by decreasing value of $C$ (Eq. (5)), followed by negative documents ranked by increasing value of $C$.

Another interesting feature of the plot is the easy identification of the *point of no threshold adjustment*. When $n_\pi(d) = \beta_\pi$ and $n_\nu(d) = \beta_\nu$ for all $\pi$ and $\nu$ entity count features, $y(d) = 1 \Leftrightarrow T(d) = \lambda - 1$ (see Figure 4, right). This means that NER information is neutral and the decision ($x(d) > \lambda - 1$) is exclusively made by the value of $x(d)$ computed from textual features via Eq. (3).

Notice that the value of $x(d)$ in Eq. (3) can be undetermined if $N(d) = 0$. Therefore, if $P(d) = N(d) = 0$, which means there is no information from textual features about document $d$ (no textual feature occurs in $d$), we compute $x(d) = \lambda - 1$, which means that decision is exclusively made by NER information. Additionally, if $P(d) > 0 \wedge N(d) = 0$, we compute $x(d) = (\lambda - 1).P(d)$, which means that the decision is made by using NER information as well as the contributions from textual features for a positive decision.

**Experimental Setting: Training and submissions**

Training of the VTT classifier consisted of exhaustively searching the parameter space that defines its linear surface, while doing (non-stratified) k-fold cross-validation (with k=4) on the training data. In this training scenario, textual features are computed from 75% of the documents and parameter search and validation is performed on the remaining 25%, for each of the cross validation runs. The parameter space is defined by $\lambda, \beta_\pi, \beta_\nu$ where $\pi \in \{1 \ldots EP\}$ and $\nu \in \{1 \ldots EN\}$. For each set of parameter values, we compute performance as the *rank product* [13, 5] of the means of the *Balanced F-Score* ($F_1$), *Accuracy*, and *Matthew's Correlation Coefficient* (MCC) measures for the 4 cross-validation folds[5]. The search is performed as follows:

1. Set all $\beta_\pi$ to the values that maximize $|p_P(n_\pi(d) \geq x) - p_N(n_\pi(d) \geq x)|$, as observed in entity count feature charts (see above). Same for $\beta_\nu$.
2. Search $\lambda, \beta_\pi, \beta_\nu$ space with coarser steps around values set in 1. Search λ widely.
3. Collect the most common values of $\lambda, \beta_\pi, \beta_\nu$ in the top echelon of classifier parameter sets obtained by the rank product of performance measures. All classifiers in the top echelon have the same value of rank product.
4. Search more finely around values obtained in 3.
5. Repeat 3 and 4 until the top echelon of classifier parameter sets is very small and one classifier can be selected with higher value of *Precision*[6].

This search procedure rewards not only the top performing classifiers, but also those parameter ranges whose performance is *robust* to small changes in the other parameters. This is achieved in step 3 of the search procedure, when we select the most common values of parameters in the (initially large) set of top performing classifiers. Because VTT is very simple to compute, the search can be done in a pretty exhaustive manner, depending on the number of parameters needed for entity count features[7].

We set out to investigate (1) if additional NER information can improve PPI article classification, (2) if there is a performance cost to using SP instead of bigram word-pair features, and (3) if the addition of full text information improves classification. To answer these questions, we submitted different versions of the VTT algorithm described below.

**No NER Information, VTT⁰:** This version uses no NER information at all, only textual features. Its decision surface is obtained simply by making $n(d) = \beta$ for all $\pi$ and ν in

---

[5] Accuracy = $\frac{TP+TN}{TP+FP+TN+FN}$, $F_1 = \frac{2.TP}{2TP+FP+FN}$, $MCC = \frac{TP.TN-FP.FN}{\sqrt{(TP+FP)(TP+FN)(TN+FP)(TN+FN)}}$, where $TP, TN, FP, FN$ refer to true positives, true negatives, false positives, and false negatives, respectively.

[6] Precision = $\frac{TP}{TP+FP}$.

[7] We provide Excel worksheet demos of the VTT surfaces and parameter search code in supplementary materials: http://cnets.indiana.edu/groups/casci/piare. These simple demos are capable of searching the entire space of BC3 data, which highlights how computationally simple the classifier is.

Eq. (3) (point of neutral NER information for every possible entity count feature in every document). This results in the simple expression below for a constant λ:

$$x(d) \geq \lambda, \qquad x(d) = \frac{P(d)}{N(d)}$$

Eq. (6)

The decision is solely defined by the sums of the (cosine and sine) contributions from the textual features for document *d*. We submitted two variations of this classifier: one computed with SP features and the other with bigrams. Since this VTT version only uses textual features extracted from titles and abstracts, these two classifiers do not use any data from the full-text documents in $D^{PMC}$ (see feature extraction above).

**ABNER Protein mentions in abstracts, VTT[1]:** This is the same classifier we used in BC2 and BC2.5 [1, 13]. In addition to textual features, it uses a single entity feature count: *ABNER protein mentions in abstracts*, which is positively correlated with positive documents. In this case, in equations (1-4), *EN*=0, *EP*=1, and *M*=1. Therefore, the decision surface (Eq. (3)) is given by:

$$x(d) \geq \lambda - y(d), \qquad x(d) = \frac{P(d)}{N(d)}, \qquad y(d) = \frac{n(d)}{\beta}$$

Eq. (7)

where β and *n(d)* refer to *ABNER protein mentions in abstracts* and λ is a constant. The initial value of β for the search algorithm (training) above is chosen as the value that maximizes the difference of occurrence probabilities of this entity count feature between the positive and the negative documents, as depicted in Figure 2: β=5. We submitted two variations of this classifier: one computed with SP features and the other with bigrams. These two classifiers also do not use any data from the full-text documents in $D^{PMC}$.

**With all NER data, VTT[5]**: This version is a substantial development from the classifier we used in BC2 and BC2.5 [1, 13], as can be seen from Eq. (3). In addition to textual features, it uses the five entity feature counts, identified earlier, that are all positively correlated with positive documents. In this case, in Equations (1-5) above, *EN*=0, *EP*=5, and *M*=5. The indices for the β and $n(d)$ values are as follows: 1 refers to *ABNER protein mentions in abstracts*, 2 refers to *NLProt protein mentions in abstracts*, 3 refers to *OSCAR compounds in abstracts*, 4 refers to *ABNER protein mentions in figure captions*, and 5 refers to *PSI-MI methods in full texts*. Therefore, its decision surface (Eq. (3)) is given by:

$$x(d) \geq \lambda - y(d), \qquad x(d) = \frac{P(d)}{N(d)}, \qquad y(d) = \frac{1}{5}\sum_{\pi=1}^{5}\frac{n_\pi(d)}{\beta_\pi}$$

Eq. (8)

where λ is a constant. Notice that because entity features 4 and 5 are extracted from full text documents, for a substantial number of documents these features do not exist in our dataset. To account for that, when a document $d$ does not have full text ($d \notin D_{PMC}$): $n_4(d) = \beta_4$ and $n_5(d) = \beta_5$, i.e. for these documents, the VTT classifier assumes the point of neutral NER information for entity features 4 and 5. The initial values of $\beta_1, \beta_2, \beta_3, \beta_4$ and $\beta_5$ for the search algorithm (training) were obtained by inspection of the charts in Figures 2 and 3, and are set to 5, 10, 15, 5, and 40, respectively. We submitted two variations of this classifier: one computed with SP features and the other with bigrams.

**With NER from abstracts only, VTT³**: this is a reduced version of VTT⁵, where we only use NER features extracted from abstracts (feature 1-3). In this case, in equations 1-5 above, *EN*=0, *EP*=3, and *M*=3. Everything else is done as for VTT⁵, using only the three entity count features from abstracts: *ABNER protein mention*, *NLProt protein mentions*, and *Oscar compound mentions*. Based on our positive experience with SP features (see results below), we only employed these in VTT³. Training was done in the exact same manner as the other classifiers, leading to the optimal parameters shown in Table 3. Its performance on training and test data is shown in Tables 4 and 5, respectively.

The final parameter values for all classifiers, obtained after the search for optimal performance on the cross-validation folds of the training data are listed in Table 3; their performance on this cross-validation is listed in Table 4. Figure 5 depicts the documents in one of the validation subsamples of the 4 cross-validation folds, and the decision surfaces of the VTT¹ and the VTT⁵ classifiers obtained with SP and bigram features.

**Results**

From our NER and dictionary tools analysis, we identified publically available resources that benefit the classification of PPI-relevant documents. Based on this analysis we selected 5 entity count features, the behavior of which for PPI classification is presented in Figures 2 and 3. Similar charts for all tools and features tested are provided in supplementary materials, including those for rejected tools. Knowledge about the behavior of these tools for PPI article classification is one of the contributions of this work.

During the challenge, our system (both online and offline) was severely hindered by various software and integration errors[8]. The various versions of the VTT classifier described above were submitted as different runs, but not at all with the correct class labels and confidence values. Therefore, the official BC3 results for our system are not only very low, but have no value with respect to the questions we set out to answer. After the challenge, we corrected all errors and computed new performance measures using the BC3 evaluation script and gold standard. Naturally, we trained the corrected classifiers without using any information from the gold standard. Demos are provided with our training (and parameter search) procedure in supplementary materials, to allow our results to be reproduced.

The performance of the corrected classifiers on the test set $D_t$ is shown in Table 5 for the *Area Under the interpolated Precison and Recall Curve* (AUCiP/R), *Balanced F-Score* ($F_1$), *Accuracy*, and MCC measures. Table 6 shows the central tendency values for these measures for all runs submitted to ACT, including our original and corrected runs. In highly unbalanced scenarios such as BC3, the accuracy measure is not as relevant or useful, since a classifier that predicts every document to belong to the dominant class will still show high accuracy. For that reason, and to provide a well-rounded assessment of performance in the unbalanced article classification scenario of Biocreative, we have proposed the use of the rank product of *AUCiP/R*, $F_1$, *MCC*, and *Accuracy* measures [13], which we refer to as RP4. Table 7 contains the performance of the top 10 submissions to ACT, as measured by RP4; Figure 6 depicts the decision surfaces for VTT[1] and VTT[5] with the documents from the test set, using SP and bigram features.

---

[8] The errors included: overwritten values of the entity count features in our database, which effectively randomized the values of these features for the test set documents; an error in the computation of the confidence measure given by Eq. (5), which tended to return the same value for most documents in the test set; and an error in the classification surface of VTT leading to many incorrect class labels.

We can see that the VTT[5] classifier performed extremely well for both versions tested (with SP and with bigrams). As can be seen in Table 7, the values of AUCiP/R, $F_1$, and MCC obtained by VTT[5] with SP features are higher than those of the top reported classifier in the challenge (team 73, Wilbur et al., Run 2) by 0.054, 0.035, and 0.037, respectively; these represent very substantial performance improvements of 8%, 5.6%, and 7.1%, respectively. The accuracy for VTT[5] was above the mean and the 95% confidence interval of the mean (see Table 6), though just below the top 20 runs for accuracy in the challenge. When evaluated by the RP4, the VTT[5] with SP features also outperforms the top reported run in the challenge. Therefore we can conclude that the VTT method, when utilizing all useful NER data, is very competitive; see analysis of results in the discussion section.

## Interaction Methods Task

### Approach and Tools

*Identifying Method Sentences:* To find candidate evidence passages in text, we used classifiers developed and reported in an earlier work by Shatkay *et al.* [22], which were trained on a corpus − unrelated to protein-protein interactions − of 10,000 sentences taken from full-text biomedical articles, and tagged at the sentence-fragment level. Each sentence in that corpus was tagged by three independent biomedical annotators, along five dimensions: *focus* (methodological, scientific or generic), *type of evidence* (experimental, reference, and a few other types), *level of confidence* (from 0 − no confidence, to 3 − absolute certainty), *polarity* (affirmative or negative statement), and *direction* (e.g. up-regulation vs. down-regulation), as described in an earlier publication [24]. The corpus itself is publically available at
*http://www.ncbi.nlm.nih.gov/pmc/artiicles/PMC2678295/bin/pcbi.1000391.s002.zip*.

While the corpus had little or nothing to do with protein-protein interaction, the Support Vector Machine (SVM) classifier (implemented using LibSVM [6]), trained along the *Focus* dimension, showed high specificity (95%), sensitivity (86%) and overall F-measure (91%) in identifying *Methods sentences*. As such, we have used it without any retraining.

Using the converted text files provided by BioCreative, we broke the text into sentences (using the Lingua-EN-Sentence Perl module [25]), and eliminated bibliographic references employing simple rules. Namely, in articles that contained a *Reference* heading, sentences following the heading were removed; when the *Reference* heading was absent, regular-expressions (based on simple patterns for identifying lists of authors, and publication dates) were used to remove likely references. The remaining sentences were represented as term vectors (as described in an earlier work [22]) and classified according to their *focus*, utilizing the SVM classifier as mentioned above, thus identifying candidate sentences that are likely to discuss methods. While we also experimented with the classifiers trained to tag text along the other dimensions, as almost all sentences were of affirmative polarity and high confidence, we decided to use only the Focus classifier; particularly, using the pertinent aspect of whether or not a sentence was classified as *a Method sentence*.

***The Methods Identifiers (MI) Dictionary:*** In order to associate the actual method identifiers with the classified sentences, we used dictionary-based pattern-matching against PSI-MI ontology terms [12]. To construct the dictionary, we obtained all the PSI-MI terms listed under the "Interaction Method" (MI:0001) branch of the ontology using the Perl module *OBO::Parser::OBOParser* [2]. The individual words within all the terms, both in the text and in the dictionary, were all stemmed using the Perl module *Lingua::Stem* [10] that implements the Porter stemmer [18]. Stemming was applied because our early experiments, without stemming, showed inferior results (data not shown). The dictionary was extended to include individual (stemmed) words occurring within the PSI-MI terms, as well as bi-grams and tri-grams of individual words occurring consecutively within the terms, produced using the Perl module *Text::NGramize* [15]. Words that are hyphen-separated within PSI-MI were included in the dictionary twice, using two forms: one in which the hyphens are replaced by spaces (thus separating the words), and another in which the hyphen is removed and the words are treated as one single composite word. The two forms allow matches against free text in which the same composition appears either completely un-hyphenated (space delimited) or collapsed into one word.

Two special cases emerged from the training set and received special treatment: (i) the tool *pdftotxt,* used by BioCreative to convert articles into plain text, consistently converted the words "*fluorescence"* into "*orescence*"; to correct for that we introduced the term *orescence* into the dictionary, as a synonym for the term *fluorescence microscopy* (MI:0416); (ii) similarly, we added the synonyms "*anti tag immunoprecipitation*" and "*anti bait immunoprecipitation*" for "*anti tag co immunoprecipitation*" (MI:0007) and "*anti bait co immunoprecipitation*" (MI:0006) respectively. These two methods are by far the most common methods identified in the training set (over 700 assignments of each, as opposed to about 480 assignments of the next popular method, MI:0096, *pull-down*). This addition ensures that occurrences within the text of the terms "anti tag immunoprecipitation" and "anti bait immunoprecipitation" constitute an exact match to MI:0006 or MI:0007 respectively, rather than an erroneous exact match to the more generic method "*immunoprecipitation*" MI:0019.

We note that while the dictionary above is based on the whole PSI-MI ontology, our final reported results consider only sentences that match terms from the reduced list of Molecular Interactions identifiers provided by BioCreative, at:
*http://www.biocreative.org/media/store/files/2010/BC3_IMT_Training.tar.gz*.

***Matching Against the Dictionary:*** Pattern matching of text against the dictionary entries was implemented using the Perl rewrite system *Text::RewriteRules* [23]. The system was customized to support both full and partial matches; to avoid a large number of spurious matches it was adjusted to prefer longer matches over shorter ones, and perfect matches over partial ones. The Perl module *Lingua::StopWords* [11] was used to avoid the matching of common English words. Sentences within which matches to the dictionary were identified, were then scored as described next.

*Scoring:* As discussed above, each sentence was tentatively associated with all the MIs whose terms (partially) matched the sentence. Statistical considerations were then used to post-process the tentative matches. When multiple MIs hit the same sentence overlapping the same word, a single MI had to be selected; similarly, a single sentence was selected as evidence for each matched MI.

We assigned a score to each sentence that was matched by an MI, based on several statistical considerations involved in associating a MI to a sentence and based on the *Focus* label assigned to the sentence, as described in the first part of this section. We first calculated an un-normalized score, which is a positive number that can be greater than 1. We normalized all scores to be between 0 and 1 as a final step.

The raw (un-normalized) score, *RScore*, for a sentence $S_i$ and a Method Identifier $MI_j$, whose dictionary entry (partly) matches the sentence, is expressed as the sum of two components:

$$\mathbf{RScore}(S_i, MI_j) = \mathbf{MIScore}(S_i, MI_j) + \mathbf{FocusScore}(S_i).$$

The first component, **MIScore**($S_i, MI_j$) is calculated based on several counts indicating how strong the association of the method identifier $MI_j$ is with the sentence $S_i$. This score is proportional to the length of the matched portion of the synonym for the MI within the sentence, measured both in characters and in words; the score is inversely proportional to the likelihood of the MI to match a sentence by chance, based on the frequency in which words from the MI synonyms occur in the dataset. To formally define the MIScore, we denote by $Hit(S_i, MI_j)$ the (partial) match of any synonym of the method $MI_j$ within sentence $S_i$, and by $|Hit(S_i, MI_j)|$ the number of *characters* within $S_i$ that actually matched the synonym. The **MIScore** itself is then calculated as the sum of the three following summands:

*Score1* rewards longer matches, but discounts such matches if they are common in the dataset:

$$Score1(Si, MIj) = 0.5 \cdot \log\left[\frac{|Hit(Si, MI_j)|}{\sum_{\text{Over all Articles } d \text{ in the dataset } D}(\text{\# of times } MI_j \text{ partially matches a sentence in } d)/|D|}\right].$$

The number of times the method identifier $MI_j$ matches a sentence in article $d$ denotes the count of any (full or partial) matches by any synonym included in the dictionary entry for $MI_j$. The term $|D|$ denotes the total number of articles in the set of articles, $D$. The log function and the multiplication by 0.5 puts Score1 in the same numerical range and order of magnitude as *Score2* and *Score3* below, and are hence employed.

*Score 2* rewards longer matches as well, but discounts such matches if the MI has typically short synonyms (as measured by the length of its individual words), and as such is more likely to have partial matches within the text by chance:

$$Score2(S_i, MI_j) = \frac{|Hit(S_i, MI_j)|}{Average\,Word\,Length\,in\,MI_j},$$

where

$$\text{Average Word Length in } MI_j = \frac{\sum_{\text{Over all synonyms } W_k \text{ of } MI_j} \# \text{ of characters in } W_k}{\sum_{\text{Over all synonyms } W_k \text{ of } MI_j} \# \text{ of words in } W_k}.$$

*Score 3* examines the ratio between the number of consecutive words constituting the match and the average number of words in the synonyms denoting *MI<sub>j</sub>*, denoted as:

$$R1 = \frac{\# \text{ of individual Words in Hit}(S_i, MI_j)}{\text{Average } \# \text{ of Words in } MI_j},$$

where

$$\text{Average } \# \text{ of Words in } MIj = \frac{\sum_{\text{Over all synonyms } W_k \text{ of } MI_j} \# \text{ of words in } W_k}{\# \text{ of synonyms of } MI_j}.$$

If the ratio *R1* is lower than 0.5, that is, if the match has fewer than half of the expected number of words denoting the method *MI<sub>j</sub>*, the match is penalized and given a score of -1; if this ratio is 1 or higher – that is, the match is much longer than the expected length of a synonym for method *MI<sub>j</sub>*, i.e. the match agrees with one of the longer synonyms for this MI – the match is rewarded with a score of 4, (which is a number in the higher range of values obtained for *Score1* or *Score2*); otherwise, the ratio R1 itself is returned (a number between 0.5 and 1). Formally:

$$Score3(Si, MIj) = \begin{cases} 4 & \text{if } 1 \leq R1; \\ R1 & \text{if } 0.5 < R1 < 1; \\ -1 & \text{if } R1 \leq 0.5. \end{cases}$$

As stated above**:**

**MIScore**($S_i$, $MI_j$) = $Score1(Si, MIj) + Score2(Si, MIj) + Score3(Si, MIj)$.

The other component of the raw score, ***FocusScore($S_i$)***, reflects the context of the matched sentence, $S_i$, that is, it accounts for the focus of the current sentence (i.e. whether it discusses a method or not) as well as for the focus of the sentences immediately preceding and following it. A sentence whose focus is *method* receives a FocusScore of at least 1. In contrast, a sentence whose focus is not *method* receives a FocusScore of 0 – unless it is followed or preceded by a *method* sentence. This reasoning takes into account the way natural language is used, which may cause the direct indications for methodology to occur within the vicinity of the sentence rather than within the sentence in-and-of-itself; thus, a bonus of 0.5 is added to the sentence's FocusScore when either the sentence before or the sentence after the current sentence is classified as a *method* sentence. Formally, for the $i^{th}$ sentence in the article, denoted $S_i$, FocusScore($Si$) is calculated as follows:

**FocusScore**($S_i$) = **IsMethod**($S_i$) + 0.5·**IsMethod**($S_{i-1}$) + 0.5·**IsMethod**($S_{i+1}$) ,

and **IsMethod**(*S*)**,** for any sentence *S,* is defined as:

$$\textbf{IsMethod}(S) = \begin{cases} 1 & \textbf{If the Focus label of } S \textbf{ is Method;} \\ 0 & \textbf{Otherwise.} \end{cases}$$

When multiple candidate MIs match a sentence while sharing some of the same words in their match, the MI who has the largest number of matched words is retained as a candidate match for the sentence. In case of a tie between two possible MIs with the same number of matching words, the MI with the longest match as measured in *characters* (rather than in words) is retained.

Finally, the evidence for a specific method $MI_j$, denoted as $Ev(MI_j)$, within an article $d$, is the sentence $S_i$ for which the raw score, **RScore**($S_i$, $MI_j$), is the highest among all other sentences within the article in which a partial match was found for a synonym of the method $MI_j$. Formally, for an article $d$, and a method identifier $MI_j$, the evidence for $MI_j$ in $d$ is: $Ev(MI_j) = \underset{Sentence\, S_i \in d}{Argmax}\, \text{RScore}(S_i, MI_j)$, and the score of this evidence is the RScore of the sentence that maximized the expression on the right hand side.

*Score Normalization:* Notably, the raw score, calculated as:

**RScore**($S_i$, $MI_j$) = **MIScore**($S_i$, $MI_j$) + **FocusScore**($S_i$) ,

is un-normalized, and as such is a positive number *not* necessarily in the range *[0,1]* as required by BioCreative. The raw scores are normalized per article, by dividing each raw score by the *maximum* raw score assigned to any *pair of method identifier and sentence* within the article. The latter step guarantees that the normalized score is always at most 1.

To produce the different runs submitted to BC3, as well as the runs described here which were produced after the workshop, the same matching and scoring algorithms were used for all runs; the difference between the different runs is merely in the threshold employed on the raw scores of evidence per method, used in order for the MI to be included or excluded in the submitted results report.

In the five runs submitted (results provided in Table S.1 of the supplementary material), Run 1 included the top 40 results for each document, while Run 5 included only methods and evidence with a raw score above 4.5 (before normalization). Unfortunately, the official runs submitted to BC3 were all produced using an erroneous code, mis-executing the pattern matching step against the dictionary and missing many valid matches. After the official submission, the errors in the code were corrected and thus the runs and the results have changed. As such, we do not provide further details on the official runs aside for reporting the official results in Table S.1, as these runs reflect a computation error rather than a methodological aspect.

The results provided in Tables 9 and 10 include four runs: One produced without any filtering, reporting all methods that partly matched each article, giving rise to a very high recall and low precision; the second reporting the top 40 scoring MIs for each article; the third reporting only MIs whose raw-score was higher than 6; and the fourth reporting only MIs whose raw-score was higher than 7. As expected, and as seen in Tables 9 and 10, the recall decreases while the precision increases with each consecutive run among these four.

***Independent Evaluation of the Results by Human Annotators:*** As our approach focused primarily on obtaining evidence for PPI-detection methods within the text, and as the BioCreative evaluation did not score this required evidence, in order to examine the quality of the evidence produced by our system, we have recruited a group of *five independent annotators*, all holding academic degrees in Biology and studying toward advanced degrees (MSs or PhD) in Molecular Biology, all proficient in the English language, and all experienced in reading and using scientific literature − particularly in areas within proteomics.

The annotators were given all the sentences produced as evidence by our system in one of our runs (the run corresponding to the third row in Table 9), a set consisting of 1049 sentences. Each sentence was independently labeled by three different annotators, each assigning one of three possible letter-tags to the sentence, indicating whether/how the sentence relates to methods for detecting protein-protein interaction (PPI). The tags were defined as follows:

> **Y** - if the sentence discusses a *method which can potentially be applied* for detecting protein-protein interaction.
> **M** - if the sentence discusses *a method, but the method is absolutely NOT* a protein-protein interaction detection method.
> **N -** if the sentence *does not discuss a method* whatsoever.

When annotators assigned the label "Y", they also had to assign a *numeric label*, indicating the actual protein-protein interaction content of the sentence, as follows:

> **2** - If protein-protein interaction (PPI) is directly and explicitly mentioned within the sentence, along with the method of detection.
> **1** - If PPI is implied in the sentence, along with the method of detection, but the PPI not explicitly stated.
> **0** - If PPI is neither implied nor mentioned in the sentence.

The sentence in the last case is not about PPI. That is, the sentence talks about a method; the method − to the best of the annotator's knowledge − has the potential to detect PPI, and hence labelled Y in the first place; but the sentence does not indicate that the method was actually applied to detecting PPI.

The inter-annotator agreement was high, as indicated by 65% of the sentences on which *all three annotators* assigned the exact same letter-tag, (a rate much higher than the 11% expected by chance, of three people assigning the same label out of three possible labels), and over 98% in which at least two annotators agreed on the letter-tag. That is, on only 17 sentences out of the 1049 there was a three-way disagreement in tag-assignment, much lower than the number expected by chance (which is about 220 sentences with total disagreement when labeling about 1000 sentences using 3 labels). The above details are provided to clarify the major characteristics of the corpus and the reliability of the annotators. Further details about this annotation effort, the corpus, and its potential utility, are beyond the scope of this paper and will be provided in a separate publication in the near future.

**Results**

We have submitted five official runs to BC3, all using the same basic strategy, varying only in the threshold of the scores applied to the data, and thus in the stringency of the filtering process. Therefore, the runs range from those favouring recall to those favouring precision. As mentioned above, the official submitted runs were produced by a version of our code that contained errors, and the resulting values were very low, both in terms of precision and in terms of recall, as well as by any other measurement. While we provide the results of these runs for the sake of completeness in the supplementary material (Table S.1), they carry no value in terms of evaluating the method described here in-and-of itself.

After fixing the error, we re-ran BC 3 evaluation script both on the training set, and over the released gold-standard test set as well. These results, as well as the results of the evaluation of one of our runs by a group of independent human annotators are discussed throughout the remainder of this section. Table 8 shows the results of running our system on the BC 3 training set, while Table 9 shows the results over the BC3 test set (the same set used for producing the results shown in Table S.1).

In both tables, the first row, labelled *All*, contains the results for a run in which all PPI detection methods that had any synonym partially-matched in any sentence, was reported as a PPI detection method relevant to the article. This run obviously has a very high recall at the cost of a very low precision. The next row (*Top 40*) shows the results from a run in which the forty top scoring MIs in each article are reported. The next two rows in both tables, report results of runs in which the criterion for including MIs was more strict, and required an un-normalized score, RScore, of at least 6 (run 3) or at least 7 (run 4).

Finally, Tables 10 and 11 summarize the basic statistics of the labels assigned by human annotators to one of the runs, namely, run 3 – the one in which the raw score required was at least 6.

## Integrating the IMT system into the ACT pipeline

We also experimented with using the output of the IMT in support of the ACT pipeline. Since our IMT system is focused on obtaining evidence for the interaction methods used, we investigated what happens to the entity count features when we crop the original document and keep only the evidence text extracted by the IMT system. That is, the entity recognition is performed not on the original text, but on the evidence portions that the IMT system outputs. We performed the same analysis of entity count features on the IMT-cropped training data. Specifically, we identified those entity count features for which $|p_P(n_\pi(d) \geq x) - p_N(n_\pi(d) \geq x)| \geq 0.3$ (see entity count feature section).

Since the IMT-cropped data contains substantially less text than the original documents, the processing time for NER and dictionary tools on the training and the test data is considerably reduced. The mean number of words per full-text article within the BioCreative corpus is 5,295.8 (Std. Dev. 1,878.6), whereas the mean number of words for an IMT-cropped document is 180.0 words (Std. Dev. 161.9). For tools such as NLProt and OSCAR, this represents more than 10 fold reduction in processing time (see supplementary material). Moreover, we observed that the characteristics of the entity count features are conserved in the IMT-cropped training data: the same 5 features emerge as positively correlated with positive documents (relevant charts are provided in supplemental materials).

This result is significant because it can save considerable computation time in future implementations of our pipeline within a curation effort.

## Discussion and Conclusion

**The Article Classification Task**

The VTT$^5$ classifier resulted in a ranking and classification performance substantially higher than all the reported submissions to the BC3 challenge, in terms of AUCiP/R, MCC, and F-Score (see results above). To address the questions raised in the beginning of this paper, we now consider the differences between the various versions of VTT. Clearly, adding the NER information improves PPI article classification. Not only is the VTT$^5$ method quite competitive when compared with all the submissions to BC3, but we can quantify the improvement in VTT performance by comparing the various versions of the method in Table 5. The AUCiP/R of VTT$^5$, with SP features, is 0.1937 higher than that of VTT$^1$, which is in turn 0.0467 higher than that of VTT$^0$. To gauge the significance of this improvement, vis a vis the variation in performance of all classifiers submitted to BC3, consider that the standard error and 95% confidence interval of the mean of AUCiP/R is 0.02 and 0.04, respectively (see Table 6). The relative performance improvement from one version of VTT to another, means that including ABNER protein mentions in abstracts alone, leads to a gain of almost 9.5%, and including the additional 4 entity count features leads to an additional gain of 35.9% in terms of the AUCiP/R measure. Therefore, the inclusion of several entity count features in VTT improved the *ranking ability* of the classifier significantly, which is what is primarily measured by AUCiP/R. The inclusion of NER information also improved substantially the *classification ability* of VTT as measured by Accuracy (VTT$^0$→VTT$^1$: 1.4% and VTT$^1$→VTT$^5$: 5.2%), F-Score (VTT$^0$→VTT$^1$: 5% and VTT$^1$→VTT$^5$: 14.4%), and MCC (VTT$^0$→VTT$^1$: 7.6% and VTT$^1$→VTT$^5$: 20.1%), the latter being the measure best suited for unbalanced scenarios. The performance of each version of the VTT, as reported in Table 5, can be contrasted to the central tendency and variation of the performance of all classifiers in Table 6. The improvement in terms of the rank product for all submissions to the ACT is also worthy of notice: out of 58 runs, VTT$^0$ was the 38$^{th}$ best classifier, VTT$^1$ was the 24$^{th}$ best, and VTT$^5$ was the best classifier. According to every performance measure, the largest improvement comes from including all of the entity count features. Therefore, there was much to gain by adding information from NLProt, PSI-MI, and OSCAR in addition to information from ABNER.

Regarding the textual features used, it is also quite clear from our results that using bigram textual features leads to worse performance than using the computationally less demanding SP features. We can see in Table 5 that for every version of VTT used, the SP features always outperformed bigrams for the AUCiP/R, $F_1$, and MCC measures. The exception is when it comes to the Accuracy obtained for VTT$^0$ and VTT$^1$; in these cases, the accuracy was larger when using bigrams. But since accuracy is not as informative in unbalanced scenarios, and because the accuracy of the top performing VTT$^5$ classifier was larger when using SP features, we can conclude that SP features lead to a better performance than bigrams. This suggests that SP features, by using only constituent words with high *S* score (see textual feature selection section), generalize the concept of PPI more effectively than bigrams. We conclude that not only is the use of the small set of SP features much more computationally efficient, it also leads to better performance of the VTT classifier.

Since two of the entity count features used on the best VTT classifier are derived from full-text data when available (via PubMed Central), i.e. based on ABNER protein mentions in figure captions (feature 4) and on PSI-MI methods in the full text (feature 5), we can conclude that full-text is at least partially responsible for the excellent performance reached by this classifier. However, as full text data was only available for 60% of the documents in the test set (see data and feature extraction section), it cannot be fully responsible for the performance improvement. To further examine this point, we computed a version of the classifier, $VTT^3$, that does not utilize these two entity features. While the performance of $VTT^3$ in the training data is just slightly lower than $VTT^5$ (see Table 4), on test data it is noticeably lower (see Table 5). We observe that inclusion of the full text features lead to approximately a 3% improvement in all performance measures. In comparison to all reported classifiers, $VTT^3$ is below the top two classifiers reported by team 73 (lead by W. John Wilbur at NCBI, Runs 2 and 4) as well as both the SP and bigram versions of $VTT^5$. Therefore, we conclude that the inclusion of data from full-text documents, even if available for little more than half of the documents in the training and test corpora, was useful and indeed contributed to obtaining the top reported classification and ranking system.

Besides its very competitive performance, the VTT classifier (in all versions tested) is defined by a simple linear surface that can be interpreted. Indeed, we can look at the parameters of table 3 (obtained via the training algorithm) and discern a "rule" of what constitutes a PPI-relevant document. We only uncovered 5 entity features positively correlated with positive documents (see entity count features section), therefore confidence in PPI-relevance increases linearly with all those features. Looking at the specific β parameter values in table 3 for $VTT^5$, we can discern a rule that states: "a document with a few ABNER protein mentions, many NLPROT protein and CHEBI chemical compound mentions in the abstract, a few ABNER protein mentions in figure captions, and many PSI-MI method mentions in the full-text tends to be PPI-relevant". The exact rule is of course defined by the VTT surface equation, but its linear nature allows us to discern the type of (vague) linguistic "rule-of-thumb" above, which is nonetheless meaningful. It is interesting to notice that the same rule emerges for both SP and bigram features.

**The Interaction Method Task**

For the IMT, the results shown in Tables 8 and 9 demonstrate that employing the scores, as shown in the three bottom rows of each table, leads to higher precision and lower recall than simply employing pattern matching (the first, *All* run in both tables). This suggests that the scoring scheme proposed helps to focus attention on sentences that are likely to contain PPI detection methods, although the resulting performance as measured by BioCreative is still low.

However, the advantage of our method remains in providing clear evidence for each decision. As BioCreative did not examine the evidence that was required from and provided by the different tools, we focused much of our efforts after the BioCreative workshop to better understand and evaluate the evidence we produce. We did this by recruiting five independent annotators with expertise in molecular biology to evaluate the results, assigning each evidence sentence to three independent annotators, who labelled our sentences, indicating their relevance to PPI detection methods, as summarised in Tables 10 and 11.

Notably, there is some discrepancy between the BioCreative evaluation and the values assigned to the results by our group of human annotators. According to the BC3 formal evaluation, as shown in Table 9, the precision of the third run (*RScore* ≥6) is about 26%. In contrast, as shown by Table 10, annotators who are also familiar with PPI detection methods and who read the sentences, deemed about 70% of the evidence for MIs produced by our system as discussing methods that are applicable to PPI detection. Moreover, as Table 11 shows, the annotators viewed about 35% (counts for Y1 and Y2 combined) of the sentences produced by our system to contain evidence that the methods were indeed applied toward the detection of PPI. In more than half of those (Y2, 19% of the total) the interacting proteins could be detected by the annotators, while in the remaining (Y1, 16% of the total) the interacting proteins were implicit rather than explicitly stated – but interaction detection through the application of the indicated method was still discussed. The above variability highlights the complexity and the possible ambiguity involved in the definition, the interpretation, and the evaluation of the IMT task.

A closer examination of individual sentences further demonstrates these differences in interpretation and evaluation of the task. Below are examples of evidence sentences that our system produces, found in articles that the BC3 gold standard judges as False Positive, but who appear to discuss PPI along with the method to detect it. The examples are formatted using the BC3 requested format, showing (in the required order, from left to right), the PubMed identifier of the article, the MI associated with it, along with the rank in the list (4, 6 and 4 in the three examples below) and the confidence score (the floating point number), followed by the evidence sentence itself:

*19224861    MI:0096 4    0.865173475604312    We found that PEDF was pulled down with Ni-NTA beads when the binding reactions included His-tagged LR or His-tagged LR90 (Fig. 2G).*

*18806265    MI:0114 6    0.620645021811025    Previous x-ray crystallography analyses suggest that CARD-CARD interactions occur via interaction between the 2 3 helical face, and the 1 4 helical face (50).*

*18819921    MI:0663 4    0.79176182685558    Using confocal microscopy, we show that trapping mutants of both PTP1B and the endoplasmic reticulum targeted TCPTP isoform, TC48, colocalize with Met and that activation of Met enables the nuclear-localized isoform of TCPTP, TC45, to exit the nucleus.*

These examples demonstrate the complexity in the task definition and in its evaluation criteria. The first example appears to be a description of experimental results observed by the authors. In contrast, the second of the three example sentences refers to a "*Previous*" experiment and provides a reference *"(50)"*. Curators whose explicit task is defined as finding only *novel* experimental evidence may view the sentence as not useful – because the evidence is not new; this is likely to be the reason why this method was not assigned to the document within the BC3 gold standard. However, these same curators can still use this sentence to back-trace the reference and recover the evidence from the original referenced paper (50). Furthermore, curators and scientists that are tasked with identifying *all* the evidence in support of an interaction, without the requirement for novelty, will still view the sentence as relevant evidence for the interaction. Notably, the BC3 IMT did not require novelty of evidence as part of the task specification. The third sentence primarily discusses the detection of co-localized proteins rather than of a direct interaction; as such it can be viewed by some curators as relevant and by others as irrelevant.

To summarize, while the utility of each specific sentence, as shown in the example above, may depend on the exact definition of the curation task, automatically identifying and highlighting such sentences can significantly narrow down the amount of text that a curator needs to examine. The above three examples all help to demonstrate the value of our method in identifying evidence sentences that are likely to be useful.

As a last point, we note that the time required for running our pipeline is realistic within a curation effort. For instance, for processing the test set of about 300 full text documents, the complete processing time was about 28 minutes (an average processing rate of over 10 documents per minute), of which about 12 minutes were consumed by the classification of each sentence along the various dimensions (*Focus, Evidence* etc.) by the multi-dimensional classifiers [22]. Most of the steps, including the classification of the sentences, can be readily performed off-line and parallelized to process multiple sentences simultaneously. Thus, the ideas presented here can be readily incorporated into an effective and useful curation pipeline.

## Authors' contributions

**AL**: Setting up and running of NER and dictionary tools, managing interaction between ACT and IMT sub-teams, constructing the IMT pipeline, devising and implementing scoring metrics for matches, running experiments for both subtasks, and participating in manuscript preparation. **MC**: Design of data storage and management for the ACT, running of ABNER, computation and integration of NER results and feature analysis, retrieval of full text data from PubMed Central, management of training and test corpora. **AW**: Setting up the Focus-classifiers as stand-alone classification tools, constructing the IMT pipeline, devising and implementing scoring metrics for matches, running experiments, integration with ACT pipeline, and participating in manuscript preparation. **AN**: computation and visualization of textual features from training and test corpora. **FP:** Design and implementation of the multi-dimensional classifiers used for the IMT task. **HS**: Conception and design of the multi-dimensional classifiers, development of the IMT pipeline and its integration with ACT, devising scoring metrics for matches, manuscript preparation. **LR**: Development and computation of VTT classifier, design and management of ACT pipeline, integration with IMT pipeline, manuscript preparation**.**

## Acknowledgements

We thank the annotators from Sharon Regan's lab and the department of Biology at Queen's Unviersity: Kyle Bender, Daniel Frank, Kyle Laursen, Brendan O'Leary and Hernan Del Vecchio. Their work as well as that of Andrew Wong's was supported by HS's NSERC Discovery and Discovery Accelerator awards #298292-08 and CFI New Opportunities Award 10437. Michael Conover and Azadeh Nematzadeh were supported with a grant from the FLAD Computational Biology Collaboratorium at the Instituto Gulbenkian de Ciencia in Oeiras, Portugal. We also thank support from these grants for travel, hosting and providing facilities used to conduct part of this research. We thank Artemy Kolchinsky for assistance in setting up the online server for the ACT.

## Tables

**Table 1. Top 10 SP features ranked with the *S* score.**

| $w_i, w_j$ | $p_P$ | $p_N$ | $S$ |
|---|---|---|---|
| interact--with | 0.3220 | 0.0442 | 0.279 |
| interact--between | 0.1071 | 0.026 | 0.081 |
| complex--with | 0.0920 | 0.0153 | 0.0768 |
| protein--interact | 0.0666 | 0.006 | 0.0606 |
| crystal--structur | 0.0804 | 0.022 | 0.0584 |
| yeast--two-hybrid | 0.0542 | 0.0 | 0.0542 |
| with--protein | 0.0619 | 0.0123 | 0.0496 |
| protein--kinas | 0.0705 | 0.0233 | 0.0472 |
| here--report | 0.086 | 0.039 | 0.047 |
| transcript--factor | 0.0856 | 0.0417 | 0.0438 |

**Table 2. Top 10 bigram features ranked with the *S* score.**

| $w_i, w_j$ | $p_P$ | $p_N$ | $S$ |
|---|---|---|---|
| interact--with | 0.3001 | 0.0397 | 0.2604 |
| interact--between | 0.1062 | 0.026 | 0.0802 |
| complex--with | 0.089 | 0.013 | 0.076 |
| crystal--structur | 0.0804 | 0.0218 | 0.0586 |
| yeast--two-hybrid | 0.0542 | 0.0 | 0.0542 |
| protein--interact | 0.052 | 0.0045 | 0.0475 |
| here--report | 0.0856 | 0.0384 | 0.0472 |
| protein--kinas | 0.0679 | 0.0224 | 0.0455 |
| transcript--factor | 0.0851 | 0.0415 | 0.0436 |
| ubiquitin--ligas | 0.0396 | 0.0031 | 0.0364 |

**Table 3. Parameter values for submitted classifiers after parameter search.**

| Classifier | Features | $\lambda$ | $\beta_1$ | $\beta_2$ | $\beta_3$ | $\beta_4$ | $\beta_5$ |
|---|---|---|---|---|---|---|---|
| $VTT^0$ | SP | 1.1 | - | - | - | - | - |
| $VTT^0$ | Bigrams | 1.1 | - | - | - | - | - |
| $VTT^1$ | SP | 1.3 | 40 | - | - | - | - |
| $VTT^1$ | Bigrams | 1.5 | 20 | - | - | - | - |
| $VTT^5$ | SP | 2.2 | 6 | 50 | 70 | 4 | 40 |
| $VTT^5$ | Bigrams | 2.1 | 6 | 50 | 60 | 5 | 30 |
| $VTT^3$ | SP | 1.4 | 17 | 115 | 115 | - | - |

**Table 4. Performance of submitted classifiers on training data.**

Shown are the mean values obtained in cross-validation by the F-Score, Accuracy, and Matthew's Correlation Coefficient. Shaded values represent best performance in table.

| Classifier | Features | $F_1$ | Accuracy | MCC |
|---|---|---|---|---|
| $VTT^0$ | SP | .7637 | .8308 | 0.6325 |
| $VTT^0$ | Bigrams | .7541 | .832 | 0.6269 |
| $VTT^1$ | SP | .7755 | .8386 | 0.6502 |
| $VTT^1$ | Bigrams | .7568 | .8302 | 0.6265 |
| $VTT^5$ | SP | .7762 | .848 | 0.662 |
| $VTT^5$ | Bigrams | .7751 | .842 | 0.6533 |
| $VTT^3$ | SP | .771 | .8387 | 0.6466 |

**Table 5. Performance of submitted classifiers on test data.**

Shown are the values obtained on the official BC3 gold standard by the F-Score, Accuracy, Matthew's Correlation Coefficient, and Area Under the interpolated Precision and Recall Curve (computed with the official script, and adding F-Score). Shaded values represent best performance in table.

| Classifier | Features | $F_1$ | Accuracy | MCC | AUCiP/R |
|---|---|---|---|---|---|
| $VTT^0$ | SP | .5399 | .8097 | .456 | .4935 |
| $VTT^0$ | Bigrams | .5243 | .8382 | .4318 | .4287 |
| $VTT^1$ | SP | .5667 | .8213 | .4909 | .5402 |
| $VTT^1$ | Bigrams | .5575 | .8402 | .472 | .5015 |
| $VTT^5$ | SP | .6483 | .864 | .5897 | .7339 |
| $VTT^5$ | Bigrams | .6366 | .85.9 | .5752 | .7127 |
| $VTT^3$ | SP | .628 | .8387 | .5735 | .7143 |

**Table 6. Central tendency and variation of the performance of all runs submitted to ACT on the official BC3 gold standard, including our original and our corrected runs.**

Shown are the values obtained by the F-Score, Accuracy, Matthew's Correlation Coefficient, and Area Under the interpolated Precision and Recall Curve (computed with the official script, adding F-Score),

|               | Accuracy | $F_1$ | MCC   | AUCiP/R |
|---------------|----------|-------|-------|---------|
| Mean          | .7909    | .4624 | .3885 | .5048   |
| Median        | .8452    | .5399 | .4608 | .5367   |
| St. dv.       | .1324    | .1732 | .1740 | .1505   |
| Mean + 95% CI | .8257    | .5079 | .4343 | .5444   |
| St. error     | .0174    | .0227 | .0229 | .0198   |

**Table 7. Performance of top 10 reported runs to ACT in BC3.**

Shown are the values obtained on the official BC3 gold standard by the F-Score, Accuracy, Matthew's Correlation Coefficient, and Area Under the interpolated Precision and Recall Curve (computed with the official script, adding F-Score), as well as their ranks. RP4 denotes the rank product of these 4 measures. Shaded values represent best and second-best performance for respective measure.

| Team | Run     | Acc.  | Rank | $F_1$ | Rank | MCC    | Rank | AUCiP/R | Rank | RP4   |
|------|---------|-------|------|-------|------|--------|------|---------|------|-------|
| **T81**  | *VTT5-SP* | .864  | 21   | .6483 | 1    | .58974 | 1    | .7339   | 1    | 21    |
| T73  | RUN_2   | .8915 | 1    | .6132 | 5    | .55306 | 4    | .6796   | 5    | 100   |
| T73  | RUN_4   | .8888 | 3    | .6142 | 4    | .55054 | 5    | .6798   | 4    | 240   |
| **T81**  | *VTT5-Bi* | .859  | 25   | .6366 | 2    | .57523 | 2    | .7127   | 3    | 300   |
| **T81**  | *VTT3-SP* | .844  | 30   | .6280 | 3    | .57345 | 3    | .7143   | 2    | 540   |
| T73  | RUN-1   | .8755 | 16   | .6083 | 6    | .53524 | 6    | .6591   | 6    | 3456  |
| T73  | RUN_3   | .8778 | 13   | .6014 | 9    | .52932 | 8    | .6589   | 7    | 6552  |
| T73  | RUN_5   | .8762 | 15   | .6033 | 8    | .53031 | 7    | .6537   | 8    | 6720  |
| T90  | RUN_3   | .8832 | 9    | .5964 | 11   | .52914 | 9    | .6524   | 9    | 8019  |
| T65  | RUN_2   | .8793 | 12   | .5982 | 10   | .52727 | 10   | .6389   | 10   | 12000 |

**Table 8. IMT Runs on the training set (after code correction)**

| Run | Precision | Recall | F-Score | MCC | AUC iP/R | Total Docs Evaluated |
|---|---|---|---|---|---|---|
| All | 2.38% | 94.80% | 0.0465 | 0.0937 | 0.2032 | 2002 |
| Top 40 | 4.54% | 85.16% | 0.0864 | 0.1598 | 0.2063 | 2002 |
| RScore ≥6 | 26.30% | 58.72% | 0.3633 | 0.3806 | 0.1997 | 1947 |
| RScore ≥7 | 29.14% | 50.25% | 0.3689 | 0.3711 | 0.1816 | 1871 |

The table shows the results of running our (corrected) program on the BC 3 training set. The measurements shown are of precision, recall, F-score, Matthews Correlation Coefficient (MCC), Area under the Curve, and the total number of articles being evaluated by our program.

The rows reflect four different runs: The first based on pattern-matching of methods to the text alone (All); the second scoring the sentence-method associations and reporting the top 40 scoring methods; the third reporting the top scoring methods whose raw score was at least 6, while the last reporting the top scoring methods whose top score was at least 7.

**Table 9. Runs on the test set (after code correction)**

| Run | Precision | Recall | F-Score | MCC | AUC iP/R | Total Docs Evaluated |
|---|---|---|---|---|---|---|
| All | 2.50% | 93.17% | 0.0487 | 0.0908 | 0.1852 | 222 |
| Top 40 | 4.83% | 82.92% | 0.0913 | 0.1604 | 0.1583 | 222 |
| RScore ≥6 | 26.61% | 50.58% | 0.3488 | 0.3535 | 0.1522 | 214 |
| RScore ≥7 | 28.44% | 48.62% | 0.3589 | 0.3591 | 0.1524 | 210 |

The table shows the results of running our (corrected) program, on the BC 3 test set. The measurements shown are of precision, recall, F-score, Matthews Correlation Coefficient (MCC), Area under the Curve, and the total number of articles being evaluated by our program.

The rows reflect four different runs: The first based on pattern-matching of methods to the text alone (All); the second scoring the sentence-method associations and reporting the top 40 scoring methods; the third reporting the top scoring methods whose raw score was at least 6, while the last reporting the top scoring methods whose top score was at least 7.

**Table 10. Summary of Evaluation by Three Human Annotators, over 1049 Evidence Sentences for PPI Methods.**

| Label | # of sentences tagged by the Majority as Label | % of sentences tagged by the Majority as Label |
|---|---|---|
| Y | 755 | 72% |
| M | 112 | 11% |
| N | 165 | 16% |

The table shows the statistics of majority annotation labelling 1049 sentences, each by three independent annotators. For each annotation value, shown in the right column, we list how many sentences were labelled with this value by at least two of the three annotators.

The possible labels are: **Y** - if the sentence discusses a method which can Potentially be applied for detecting protein-protein interaction; **M** - if the sentence discusses a method, but the method is NOT a protein-protein interaction detection method; **N** - if the sentence DOES NOT discuss a method.

Note that the total number of majority-vote sentences is 1032 rather than 1049, because on 17 sentences the 3 annotators had a 3-way disagreement. (Roughly 1% of the sentences, hence the total percentage is 99%)

**Table 11. The Distribution of the Secondary Labels for Sentences Tagged as Y by Majority of Annotators**

| Label | # of sentences tagged by the Majority as Label | % with respect to all Y-tagged sentences (755) | % with respect to all sentences (1049) |
|---|---|---|---|
| Y2 | 199 | 26% | 19% |
| Y1 | 172 | 23% | 16% |
| Y0 | 297 | 39% | 28% |

Annotators assigning a "Y" to a sentence were further asked to assign a numeric label, indicating the actual protein-protein interaction content of the sentence, as follows: **2** - If Protein-protein interaction (PPI) is directly and explicitly mentioned within the sentence (along with the method of detection); **1** - if PPI is implied in the sentence (along with the method of detection), but not explicitly stated; **0** - if PPI is neither implied nor mentioned in the sentence.

The table shows the number of sentences labelled as Y2, Y1 and Y0 by a majority of the annotators, as well as the percentage with respect to the total number of sentences labelled as Y, and with respect to the whole collection of labelled sentences.

Note that the total number of majority Y2, Y1 and Y0 labels in the second column on the left does not sum to 755 (and the respective percentages do not sum to 100%), as for some of the sentences in which two or more annotators agree on the "Y" tag, there is not necessarily such agreement on the additional numerical label (0, 1 or 2).

# Figures

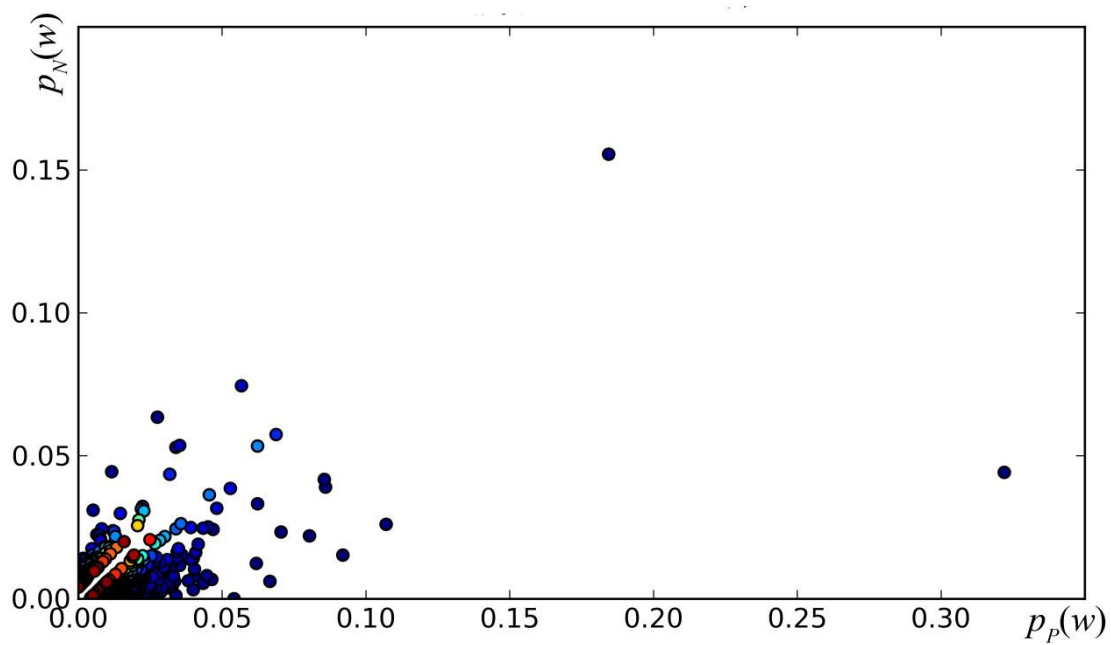

**Figure 1. Top 1000 SP Features on the $p_P/p_N$ plane. Features are colored according to the value of $S$. (darker indicating higher rank)**

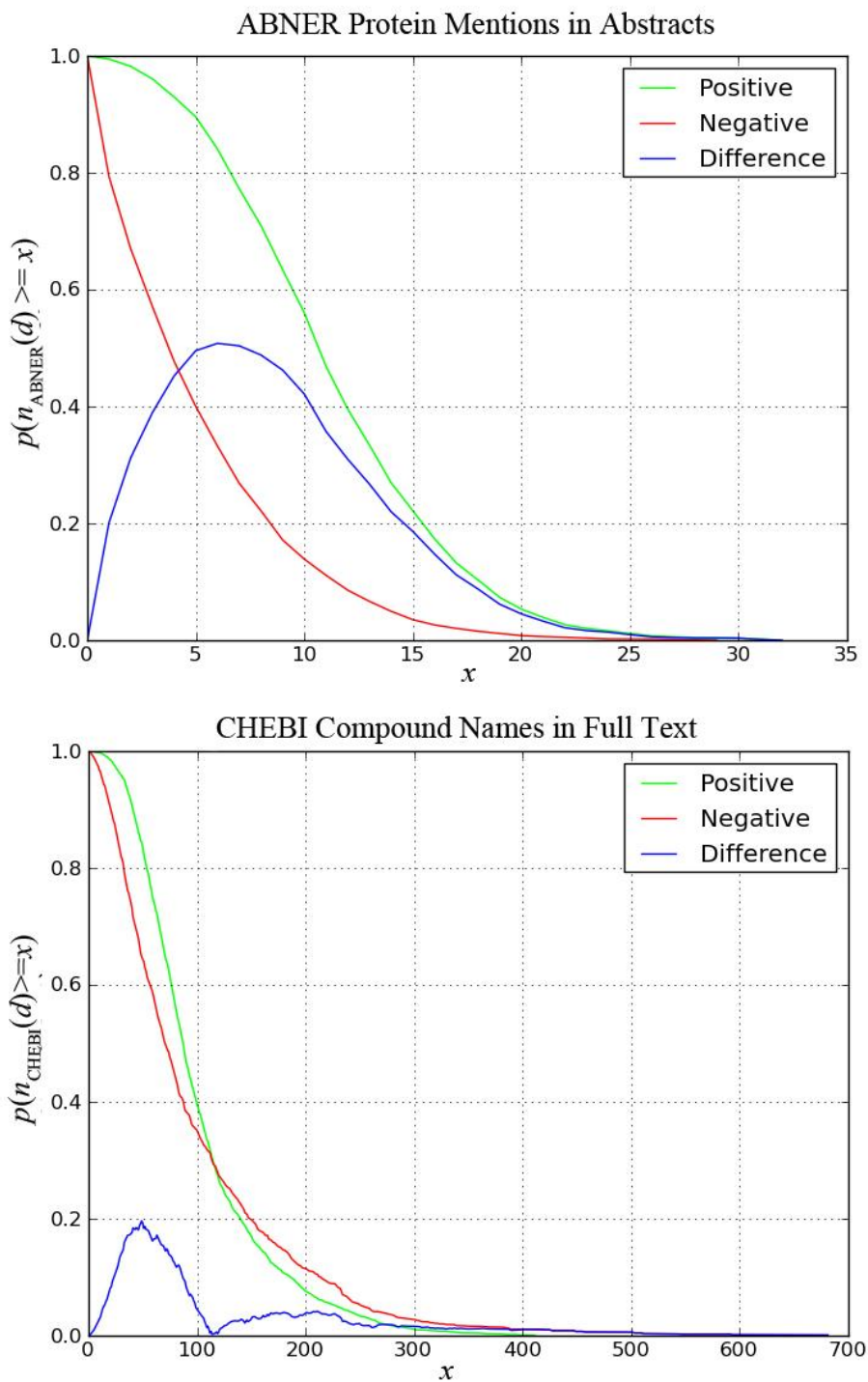

**Figure 2. Comparison of entity count features for *ABNER protein mentions in abstracts* in training set *D* (top), and *CHEBI compound names in full text* documents in training data $D^{PMC}$ (bottom). The horizontal axis represents the number of mentions $x$, and the vertical axis the probability of documents with at least $x$ mentions. The green lines denote probabilities for documents labeled relevant $p_P(n_\pi \geq x)$, while the red lines denote probabilities documents labeled irrelevant $p_N(n_\pi \geq x)$; the blue lines denote the difference between green and and red lines ($|p_P - p_N|$).**

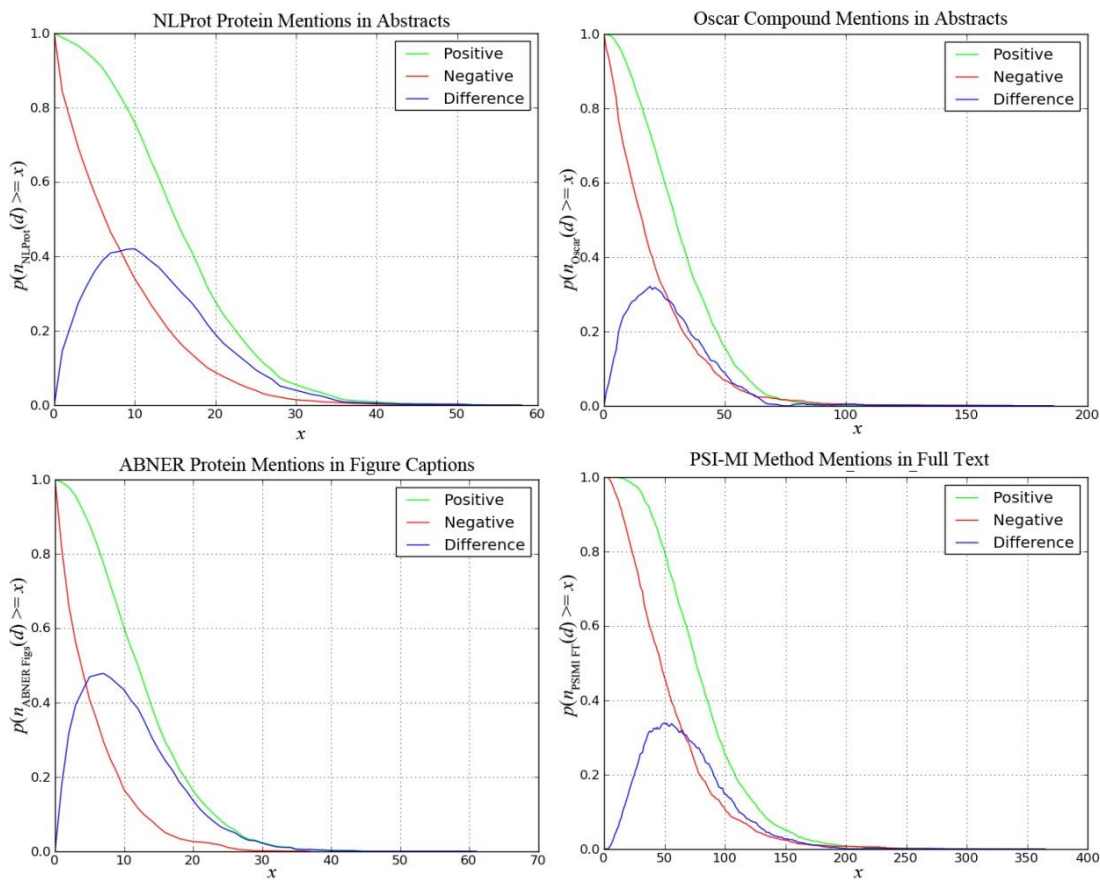

**Figure 3. Comparison of entity count features for *NLProt protein* and *OSCAR compound* mentions in abstracts in training set *D* (top), and *ABNER Protein mentions in figure captions* and *PSI-MI method mentions* in full text documents in training data $D^{PMC}$ (bottom). The horizontal axis represents the number of mentions $x$ and the vertical axis the probability of documents with at least $x$ mentions. The green line denotes probabilities for documents labeled relevant $p_P(n_\pi \geq x)$, while the red line denotes probabilities for documents labeled irrelevant $p_N(n_\pi \geq x)$; the blue line denotes the difference between green and and red lines ($|p_P - p_N|$).**

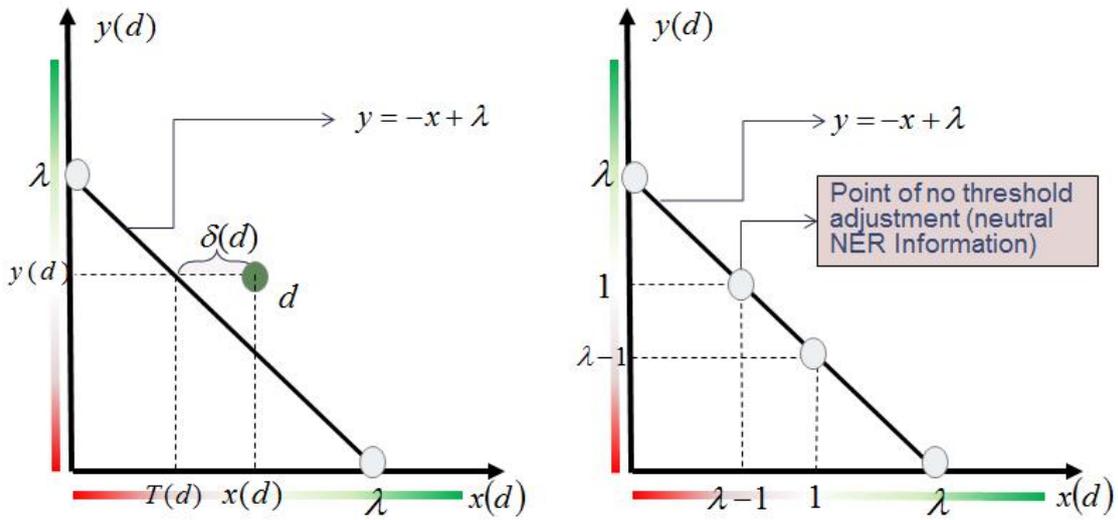

**Figure 4. The normalized plane for plotting the VTT decision surface; $x(d)$ and $y(d)$ are computed according to Eq. (3) for every document $d$. The decision surface is computed with Eq. (4). On the left-hand side the threshold for the classification decision is shown (see text for description). On the right-hand side, the point of no threshold adjustment is shown (see text for description).**

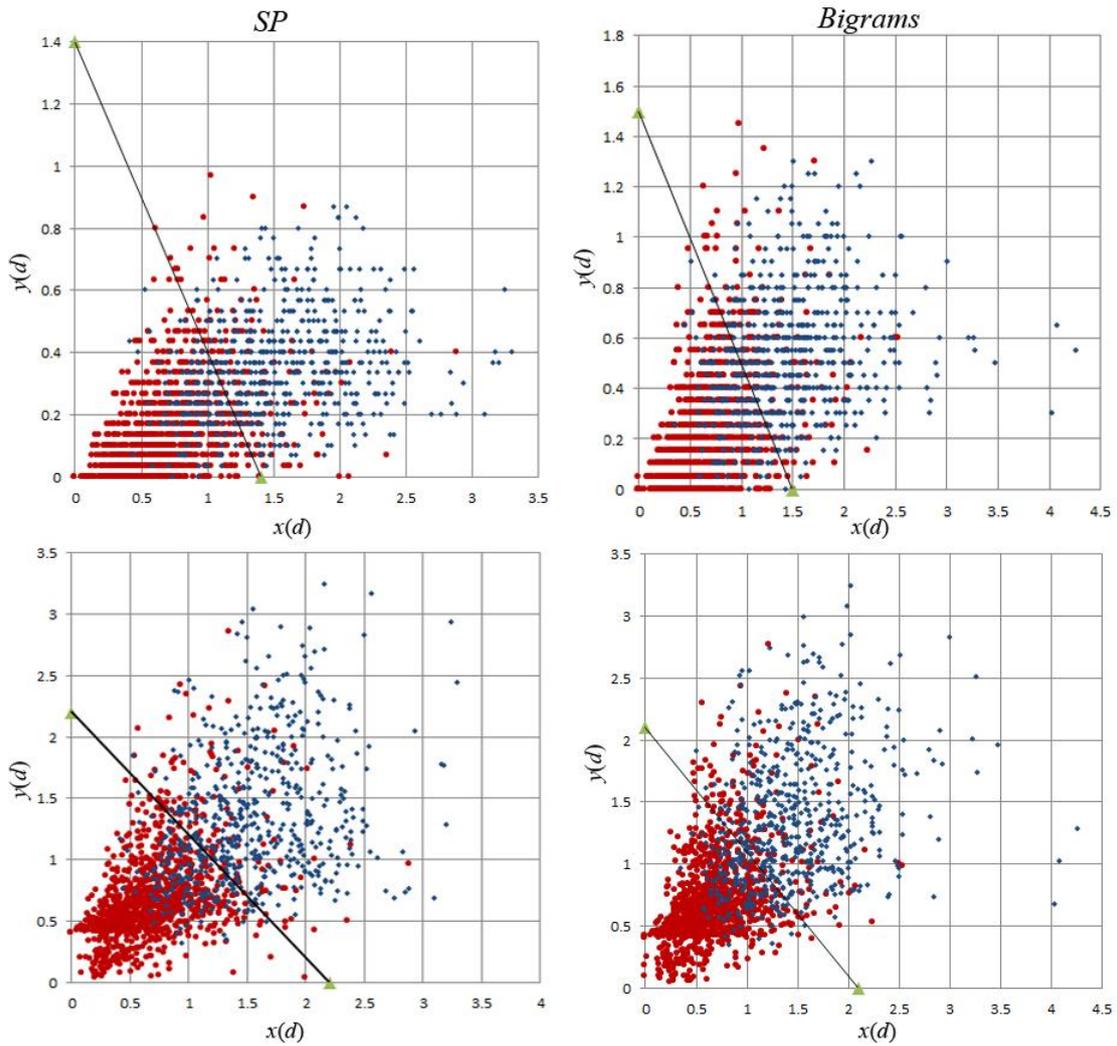

**Figure 5. Decision surfaces of the VTT[1] (top) and VTT[5] (bottom) classifiers with SP (left) and bigram (right) textual features, for the documents in one of the validation subsamples of the 4 cross-validation folds using the training data. The decision surfaces are plotted with the parameters in Table 3, and $x(d)$ and $y(d)$ are computed according to Eq. (7) for every document $d$. The plots for VTT[1] surfaces display many documents $d$ with the same values of $y(d)$, plotted in horizontal rows, while VTT[5] displays a smoother ranking of documents. This happens because VTT[1] uses information from a single NER tool (ABNER protein mentions), while VTT[5] uses information from five such tools; thus, while in the VTT[1] plot many documents have the same value of ABNER protein mentions, in the VTT[5] plot the various NER measurements lead to a finer distinction between documents.**

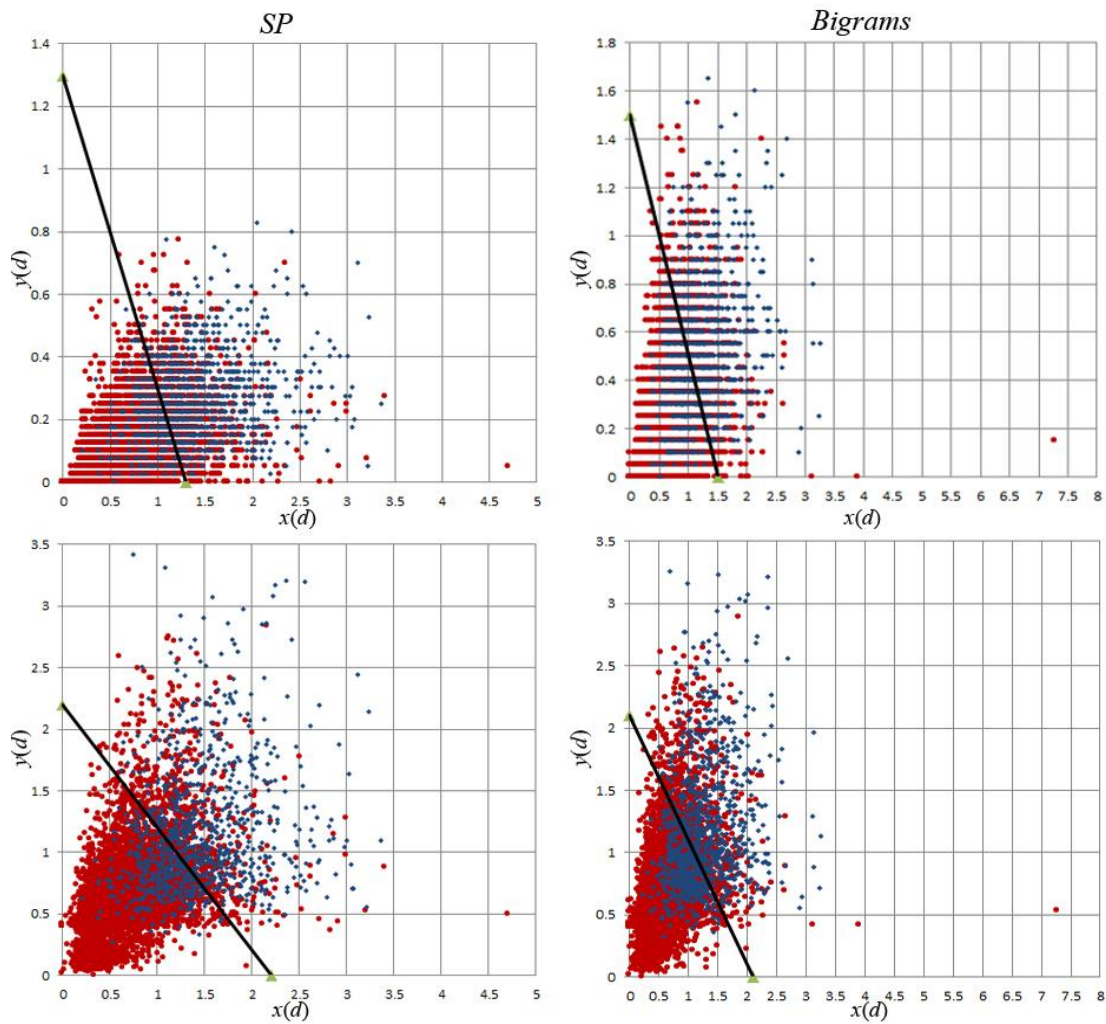

**Figure 6.** Decision surfaces of the VTT[1] (top) and VTT[5] (bottom) classifiers with SP (left) and bigram (right) features, for the documents in test data. The decision surface and $x(d)$ and $y(d)$ are computed according to Eq. (7) for VTT[1] (top) and Eq. (8) VTT[5] (bottom), for every document $d$ in test set. The plots for VTT[1] surfaces display many documents $d$ with the same values of $y(d)$, plotted in horizontal rows, while VTT[5] displays a smoother ranking of documents. This happens because VTT[1] uses information from a single NER tool (ABNER protein mentions), while VTT[5] uses information from five such tools; thus, while in the VTT[1] plot many documents have the same value of ABNER protein mentions, in the VTT[5] plot the various NER measurements lead to a finer distinction between documents.